\documentclass[runningheads]{llncs}
\usepackage{xcolor}
\usepackage{geometry}
\usepackage{graphicx}
\usepackage{url}
\usepackage{multirow}
\usepackage{array}
\usepackage{booktabs}
\usepackage{enumitem}
\usepackage{float}
\usepackage{amssymb}
\usepackage{amsmath}
\usepackage{longtable}
\usepackage{pifont}
\usepackage{color}
\usepackage{cite}
\usepackage{subfigure}
\usepackage{changepage}
\usepackage[ruled,linesnumbered]{algorithm2e}
\usepackage[misc]{ifsym}
\newcommand{\tabincell}[2]{\begin{tabular}{@{}#1@{}}#2\end{tabular}}

\begin{document}
\title{Learning to Coordinate via Multiple Graph Neural Networks}
\titlerunning{Learning to Coordinate via Multiple Graph Neural Networks}
\author{Zhiwei Xu \and Bin Zhang \and Yunpeng Bai  \and Dapeng Li \and Guoliang Fan\Letter}
\authorrunning{Zhiwei Xu, Bin Zhang, Yunpeng Bai, Dapeng Li, Guoliang Fan}
\tocauthor{Zhiwei Xu, Bin Zhang, Yunpeng Bai, Dapeng Li, Guoliang Fan}
\institute{Institute of Automation, Chinese Academy of Sciences. \\University of Chinese Academy of Sciences School of Artificial Intelligence.\\
\email{\{xuzhiwei2019, zhangbin2020, baiyunpeng2020, lidapeng2020, guoliang.fan\}@ia.ac.cn}}
\maketitle 
\begin{abstract}
The collaboration between agents has gradually become an important topic in multi-agent systems. The key is how to efficiently solve the credit assignment problems. This paper introduces MGAN for collaborative multi-agent reinforcement learning, a new algorithm that combines graph convolutional networks and value-decomposition methods. MGAN learns the representation of agents from different perspectives through multiple graph networks, and realizes the proper allocation of attention between all agents. We show the amazing ability of the graph network in representation learning by visualizing the output of the graph network, and therefore improve interpretability for the actions of each agent in the multi-agent system.

\end{abstract}

\section{Introduction}

In the past decade, multi-agent systems (MAS) have received considerable attention from researchers due to their extensive application scenarios. The change of the environment is no longer determined by a single agent but is the result of the joint actions of all agents in MAS, which results in the traditional single-agent reinforcement learning algorithm cannot be directly applied to the case of Multi-Agent. In the field of cooperative multi-agent reinforcement learning, since the dimensionality of the joint action space of multi-agents will increase exponentially as the number of agents increases, the centralized method of combining multiple agents as a single agent for training cannot achieve desired results. In addition, there is a decentralized approach represented by Independent Q-Learning (IQL)\cite{Tampuu2017MultiagentCA}, in which each agent learns independently, using other agents as part of the environment, but this method is unstable and easy to overfit. At present, centralized training and distributed execution (CTDE)\cite{Lowe2017MultiAgentAF} are the most popular learning paradigms, in which we can use and share some global information during training to make the distributed execution more effective, so as to improve learning efficiency.

On the one hand, it's better to learn a centralized action-value function to capture the effects of all agents' actions. On the other hand, such a function is difficult to learn. Even if it can be learned, there is no obvious way for us to extract decentralized policy. Facing this challenge, the COMA\cite{Foerster2018CounterfactualMP} algorithm learns a fully centralized Q-value function and uses it to guide the training of decentralized policies in an actor-critic framework. Different from this method, researchers have proposed another value-based algorithm. The main idea is to learn a centralized but decomposable value function. Both Value-Decomposition Network (VDN)\cite{Sunehag2018ValueDecompositionNF} and QMIX\cite{Rashid2018QMIXMV} adopt this idea. VDN approximates joint action-value function as the linear summation of the individual value functions obtained through local observations and actions, but in fact, the relationship between joint action-value and individual action-value is much more complicated than this, besides, VDN ignores any additional state information available during learning. The QMIX algorithm relaxes the restriction on the relationship between the whole and the individual. It approximates joint Q-value function through a neural network and decomposes it into a monotonically increasing function of all individual values. In addition, there are many excellent works in the field of value function decomposition, such as QTRAN\cite{Son2019QTRANLT} that directly learn the joint action value function and then fit residuals with another network. 

The above-mentioned value-decomposition methods have achieved good results in the SMAC\cite{Samvelyan2019TheSM} testbed. But it's worth noting that the aforementioned algorithms mainly focus on the value decomposition for credit assignment, but the underlying topology between agents in the MAS is not paid attention to or utilized. When we take this structure into account, a natural idea is to use graph structure for modeling. For data in an irregular or non-Euclidean domain, graph convolutional networks (GCNs)\cite{Defferrard2016ConvolutionalNN, Velickovic2018GraphAN, Thekumparampil2018AttentionbasedGN, Perozzi2014DeepWalkOL, Hamilton2017InductiveRL, Xu2019HowPA, Xu2018RepresentationLO} can replace traditional convolution operations and perform graph convolutions by taking the weighted average of a node's neighborhood information, so as to use the geometric structure of the graph to learn the embedding feature of each node or the whole graph. Recently, many graph convolutional networks based on different types of aggregators have been proposed, and significant results have been obtained on many tasks such as node classification or graph classification. Since the agents in the MAS can communicate and influence each other, similar to social networks, some works that combines graph networks and multi-agent reinforcement learning have appeared. Most of them can be seen as variants that increase the communication between agents. For example, CommNet\cite{Sukhbaatar2016LearningMC}, BiCNet\cite{Peng2017MultiagentBN}, and DGN\cite{Bhmer2020DeepCG} all use different convolutional kernels to process the information transmitted by neighbor agents.

In this paper, we propose a multi-agent reinforcement learning algorithm based on the CTDE structure that combines graph convolutional neural networks with value-decomposition method, namely Multi-Graph Attention Network (MGAN). We establish an undirected graph, and each agent acts as a node in the graph. Based on this graph, we build multiple graph convolutional neural networks and the attention mechanism\cite{Vaswani2017AttentionIA} is used in the aggregators. The input of the network is the individual value function obtained by a single agent, and the output of the network is the global value function. At the same time, in order to ensure that the local optimal action is the same as the global optimal action, the MGAN algorithm also satisfies the monotonicity assumption. Graph convolutional network effectively learns the vector representation of the agents in MAS, making the efficiency and accuracy of centralized training higher than other algorithms. Our experiments also show that the MGAN algorithm is superior in performance to the baseline algorithms, especially in the scenarios of a large number of agents.

\noindent
\textbf{Contribution}
\begin{itemize}
    \item We propose MGAN, a multi-agent reinforcement learning algorithm that combines graph convolutional networks and value-decomposition methods. The graph network is used to make full use of the topological structure between agents, thereby increasing the speed of training.
    \item The graph networks can learn the vector representation of each agent in the embedding space. By visualizing these vectors, we can intuitively understand that all agents are divided into several groups at each step, thereby improving interpretability for the agents’ behaviors.
    \item We demonstrate through experiments that the proposed algorithm is comparable to the baseline algorithms in the SMAC environment. In some scenarios with a large number of agents, MGAN significantly outperforms previous state-of-the-art methods.
\end{itemize}

\section{Background}
\subsection{Dec-POMDP}
A fully cooperative multi-agent task can be modeled as a decentralized partially observable Markov decision process (Dec-POMDP)\cite{Oliehoek2016ACI} in which each agent only takes a local observation of the environment. A typical Dec-POMDP can be defined by a tuple $G = <\mathcal{S},\mathcal{U}, \mathcal{P}, \mathcal{Z}, r, \mathcal{O}, n, \gamma>$. $s \in \mathcal{S}$ is the global state of the environment. At each timestep, every agent $a \in \mathcal{A} := \left \{1, ..., n\right \}$ will choose an individual action $u_a \in \mathcal{U}$. The joint action takes the form of $\boldsymbol{u} \in \boldsymbol{\mathcal{U}} \equiv \mathcal{U}^n$. $\mathcal{P}$ denotes the state transition function. All the agents in Dec-POMDP share the same global reward function $r(s, u): \mathcal{S} \times \boldsymbol{\mathcal{U}} \rightarrow \mathbb{R}$. According to the observation function $\mathcal{O}(s, a): \mathcal{S} \times \mathcal{A} \rightarrow \mathcal{Z}$, each agent $a$ gets local individual partial observation $z \in \mathcal{Z}$. $\gamma \in [0, 1)$ is the discount factor. 

In Dec-POMDP, each agent $a$ has its own action-observation history $\tau_a \in T \equiv(\mathcal{Z}\times \mathcal{U})$. The policy of each agent $a$ can be written as $\pi_a (u_a | \tau_a) : T \times \mathcal{U} \rightarrow \left[0, 1\right]$. Our aim is to maximize the discounted return $R^t = \sum_{l=0}^\infty \gamma^lr_{t+l}$. The joint action-value function can be computed by the following equation: $Q^{\boldsymbol{\pi}}(s_t, \boldsymbol{u}_t) = \mathbb{E}_{s_{t+1:\infty},\boldsymbol{u}_{t+1:\infty}}\left[R_t|s_t,\boldsymbol{u}_t\right]$, where $\boldsymbol{\pi}$ is the joint policy of all agents.

\subsection{Value-Decomposition Multi-Agent RL}
In the cooperative multi-agent reinforcement learning problem, one of the most basic solutions is to learn action-value function of each agent independently. It’s more related to the individual agent’s observations. However, previous studies indicate that this method is often very unstable and it is very difficult to design an efficient reward function. By contrast, learning the overall joint reward function is the other extreme. A key limitation of this method is that the problem of "lazy agents" often occurs, i.e., only one agent active and the other being "lazy".

To solve this issue, many researchers have proposed various methods lying between the extremes of independent Q-learning and centralized Q-learning, such as VDN, QMIX and QTRAN, which try to achieve automated learning decomposition of joint value function by the CTDE method. These value-decomposition methods are based on the Individual-Global-Max (IGM)\cite{Son2019QTRANLT} assumption that the optimality of each agent is consistent with the optimality of all agents. The equation that describes IGM is as follows:
\begin{equation*}
    \arg \max _{\boldsymbol{u}} Q_{\mathrm{tot}}(\boldsymbol{\tau}, \boldsymbol{u})=\left(\begin{array}{c}
    \arg \max _{u_{1}} Q_{1}\left(\tau_{1}, u_{1}\right) \\
    \vdots \\
    \arg \max _{u_{n}} Q_{n}\left(\tau_{n}, u_{n}\right)
    \end{array}\right),
\end{equation*}
where $Q_{tot}$ is global action-value function and $Q_a$ is the individual ones.

VDN assumes that the joint value function is linearly decomposable. Each agent learns the additive value function independently. VDN aims to learn the optimal linear value decomposition from the joint action-value function to reflect the value function of each agent. The sum $Q_{tot}$ of all individual value functions is given by
\begin{equation*}
    Q_{tot}(s,u_a) = \sum_{a=1}^n Q_a(s,u_a).
\end{equation*}
 By this method, spurious rewards can be avoided and training is easier for each agent. However, because the additivity assumption used by VDN is too simple and there are only few applicable scenarios, a nonlinear  global value function is proposed in QMIX. QMIX introduces a new type of value function module named mixing network. In order to satisfy the IGM assumption, it is assumed that the joint action-value function $Q_{tot}$ is monotonic to the individual action-value function $Q_a$:
\begin{equation*}
    \frac{\partial Q_{tot}(\boldsymbol{\tau}, \boldsymbol{u})}{\partial Q_{a}\left(\tau_{a}, u_{a}\right)} \geq 0, \quad \forall a \in \{1,\dots,n\}.
\end{equation*}

Furthermore, QTRAN uses a new approach that can relax the assumption. However, several studies have indicated that the actual performance of the QTRAN is not very good because of its relaxation.

\subsection{Graph Convolutional Networks}

Convolutional graph neural network, as a kind of graph neural network, is often used to process data of molecules, social, biological, and financial networks. Convolutional graph neural networks fall into two categories, spectral-based and spatial-based. Spectral-based methods analyze data from the perspective of graph signal processing. The spatial-based convolutional graph neural network processes the data of graph by means of information propagation. The emergence of graph convolutional network has well unified these two methods.

Let $G = (V, E)$ be a graph. Each node $v\in V$ in the graph has its own feature, which is denoted as $h_v^{(0)}$. Assuming that a graph convolutional network has a $K$-layers structure, then the hidden output of the $k$-th layer of the node $v$ is updated as follows:
\begin{equation}
\begin{aligned}
    a_v^{(k)}&=AGGREGATE^{(k)}(\{h_u^{(k-1)}|u\in \mathcal{N}(v)\}),\\
    h_v^{(k)}&= COMBINE^{(k)}(a_v^{(k)}, h_v^{(k-1)}),
\end{aligned}
\label{eq:graph}
\end{equation}
where $COMBINE$ is often a 1-layer MLP, and $\mathcal{N}$ is the neighborhood function to get immediate neighbor nodes. Each node $v \in V$ aggregates the representations of the nodes in its immediate neighborhood to get a new vector representation. With the introduction of different $AGGREGATE$ functions, various variants of the graph convolutional network have obtained desired results on some datasets. For example, in addition to the most common mean aggregators, Graph Attention Network (GAT) \cite{Velickovic2018GraphAN} uses attention aggregators and Graph Isomorphism Network (GIN) \cite{Xu2019HowPA} uses sum aggregators, both of which have achieved better results.

\section{MGAN}

\begin{figure*}[htp]
    \centering
    \includegraphics[width = 5.5 in]{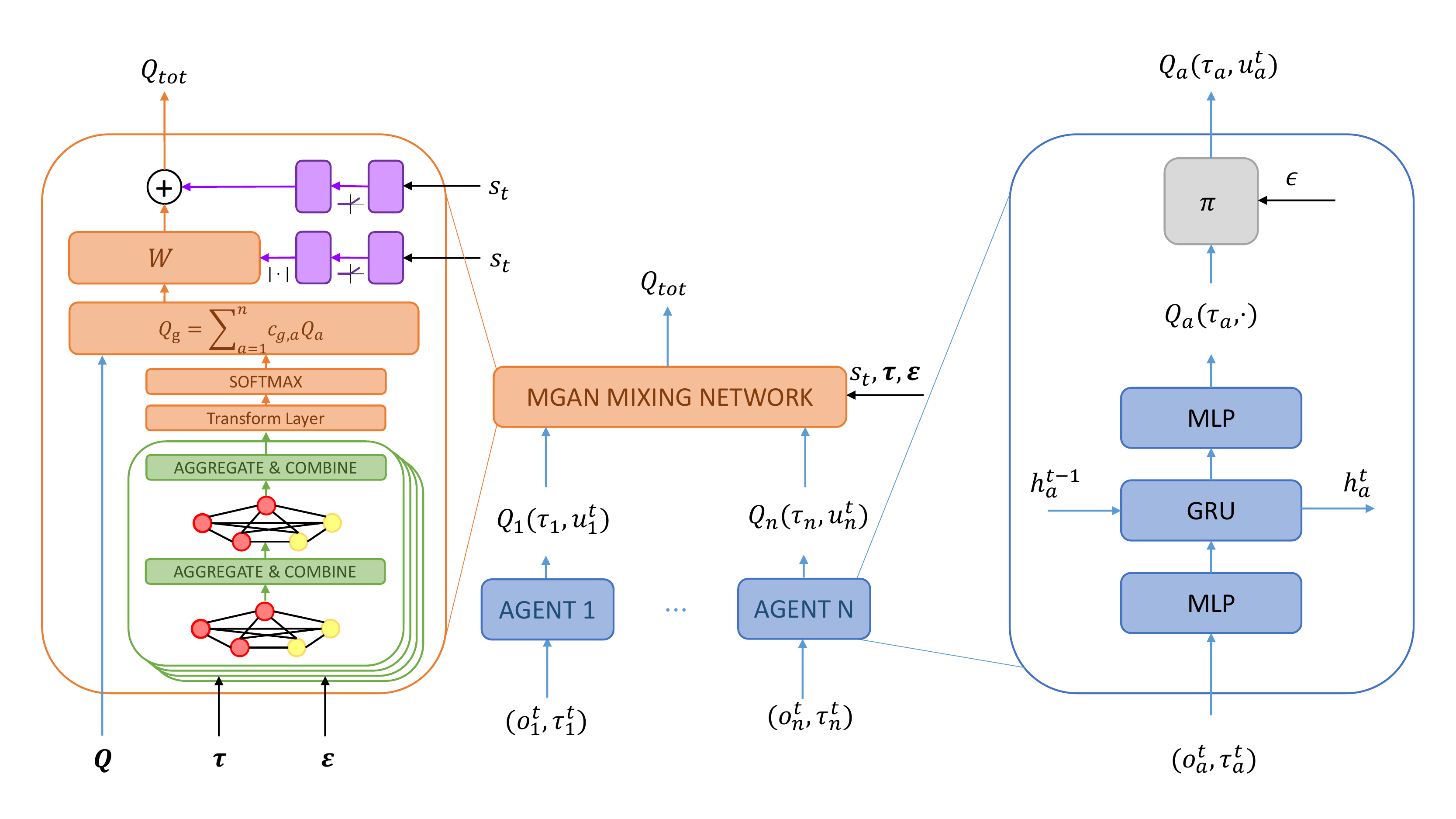}
    \caption{The overall architecture of MGAN.}
    \label{fig:Framework}
\end{figure*}

In this section, we will propose a new method called MGAN. By constructing multiple graph convolutional networks at the same time, each graph convolutional network has its own unique insights into the graphs composed of agents. This algorithm can not only make full use of the information of each agent and the connections between agents, but also improve the robustness of the performance.

\subsection{Embedding generation via graph networks}
First, we need to construct all agents as a graph $G=(V, E)$, where each agent $a$ can be seen as a node in the graph $v \in V$, i.e., agent $a$ and node $v$ has a one-to-one correspondence. We define the neighborhood function $\mathcal{N}$ to get the immediate neighbor nodes of the specified node. The edge $e_{uv}$ between any two nodes in the graph is defined as:
\begin{equation}
    e_{uv}=\begin{cases}1,&\text{if}\ u \in \mathcal{N}(v)\ \text{or}\ v \in\mathcal{N}(u)\\0,&\text{otherwise}\end{cases}
\label{eq:adjmatrix}
\end{equation}
and according to this definition, we get the adjacency matrix $\mathcal{E}\in \mathbb{R}^{n\times n}$. In the reinforcement learning tasks, the adjacency matrix often indicates whether the agents are visible or whether they can communicate with each other. Each node $v$ has its own feature $h_v$. 

Then we build a two-layer graph convolutional network to learn the embedding vector of each agent. To build a graph convolutional network, we need to define the $AGGREGATE$ and $COMBINE$ functions mentioned by Equation~(\ref{eq:graph}). Considering the actual situation, agents often need to pay special attention to a few of all other agents in the real tasks. So mean aggregators are often not qualified for this task. We adopted a simplified dot-product attention mechanism to solve this problem. The vector $a_v$ obtained by the node $v$ through the attention aggregate function can be expressed as:
\begin{equation*}
\begin{aligned}
    a_v &= AGGREGATE(\{h_u|u\in \mathcal{N}(v)\})\\
    &=\sum_{u \in \mathcal{N}(v)}\frac{\exp((h_v)^T\cdot (h_u))}{\sum_u \exp((h_v)^T \cdot (h_u))}\cdot h_u.
    \end{aligned}
\end{equation*}
Then $a_v$ needs to be entered into the $COMBINE$ function. It can be clearly seen that the embedding vectors obtained after the $AGGREGATE$ function processing loses the original characteristics of the node itself, i.e., the feature of the node is over smooth, and the characteristic information of the node itself is lacking. Therefore, we define the next layer's representation $h_v^\prime$ of the node $v$ i.e. output by the $COMBINE$ function as:
\begin{equation*}
    h_v^\prime=COMBINE(a_v)=ReLU(MLP(CONCAT(a_v, h_v)))
\end{equation*}
This step completes the nonlinear transformation of the features obtained after the node $v$ aggregates its neighbor nodes. Note that the MLP in the $COMBINE$ function of each layer is shared for each node. Similar to the simplified JK-Net\cite{Xu2018RepresentationLO}, the original feature $h_v$ is concatenated with the aggregate feature to ensure that the original node information will not be lost. From another perspective, this is very similar to ResNet\cite{He2016DeepRL}.

\subsection{MGAN Mixing Network}
Each agent corresponds to a DRQN\cite{Hausknecht2015DeepRQ} network to learn individual action-value $Q_a$, where $a\in \{1, \dots,n\}$. We have defined the graph convolutional network used to obtain the embedding vector of the agent, and then we will explain how to construct the network fitting joint action value function $Q_{tot}$. The embedding vector obtained through graph convolutional network is input into a fully connected layer neural network, which we call a transform layer, so that the embedding vector of each node $v$ is transformed into a scalar $c_v$ through affine transformation. The joint action-value function obtained by this graph convolutional network can be obtained by the following equation:
\begin{equation*}
    \sum_{a=1}^n (Q_a \cdot \frac{\exp(c_a)}{\sum_{v\in V} \exp(c_v)}),
\end{equation*}
which connects the vectors output by the graph networks with the individual action-values through dot multiplication.

Inspired by the multi-head attention mechanism, we propose to use multiple graph convolutional networks to jointly learn the embedding representation of nodes. Multiple graphs allow the model to jointly attend to information from different embedding spaces. Multiple graph convolutional networks share a transform layer. We set the number of graph convolutional networks to $G$. Thus, the following equation of the value function corresponding to each graph convolutional network is obtained:
\begin{equation*}
    Q_g=\sum_{a=1}^n (Q_a \cdot \frac{\exp(c_{g,a})}{\sum_{v\in V} \exp(c_{g,v})}), \quad \forall g\in\{1,\dots,G\}.
\end{equation*}
where $c_{g,v}$ is the scalar output by the $v$-th node in the $g$-th graph convolutional network after the transform layer.

VDN obtains the global action-value by simply summing the individual action-values of all agents. And QMIX uses multiple hypernetworks\cite{Ha2017HyperNetworks}, inputs state $s$, and outputs network weight parameters to construct a Mixing Network. It should be noted that in order to satisfy the monotonicity assumption proposed by QMIX, the network weight parameters output by hypernetworks are all positive. Our weighted linear factorization lies between the two and has a stronger representational capability for the joint value function than VDN while keeping a linear decomposition structure. This is because we only use hypernetworks to generate a layer of mixing network to linearly combine multiple $Q_g$. The entire network framework of the MGAN algorithm is shown in the Figure~\ref{fig:Framework}.

\subsection{Loss Function}
MGAN is the same as other recently proposed MARL algorithms in that they are all trained end-to-end. The loss function is set to TD-error, which is the same as the traditional value-based reinforcement learning algorithm \cite{Sutton2005ReinforcementLA}. We denote the parameters of all neural networks as $\theta$ and MGAN is trained by minimizing the following loss function:
\begin{equation*}
    \mathcal{L}(\theta) = (y_{tot}-Q_{tot}(\boldsymbol{\tau},\boldsymbol{u}|\theta))^2,
\end{equation*}
where $y_{tot}$ is the target joint action-value function and $y_(tot)=r+\gamma \max_{\boldsymbol{u^\prime}}Q_{tot}(\boldsymbol{\tau}^ \prime,\boldsymbol{u}^\prime|\theta^-)$. $\theta^-$ are the parameters of the target network.

\begin{table}[htbp]
\centering
\caption{Maps in different scenarios.}
\begin{tabular}{lcc|lcc}
\hline
Name&Ally Units&Enemy Units&$\;\;$Name&Ally Units&Enemy Units\\
\hline
& &  \\[-6pt]
2s3z&\tabincell{c}{2 Stalkers\\3 Zealots}&\tabincell{c}{2 Stalkers\\3 Zealots}&$\;\;$
3s5z&\tabincell{c}{3 Stalkers\\5 Zealots}&\tabincell{c}{3 Stalkers\\5 Zealots}\\
\hline
& & \\[-6pt]
1c3s5z&\tabincell{c}{1 Colossus\\3 Stalkers\\5 Zealots}&\tabincell{c}{1 Colossus\\3 Stalkers\\5 Zealots}&$\;\;$
8m\_vs\_9m&8 Marines&9 Marines\\
\hline
& & \\[-6pt]
2c\_vs\_64zg&2 Colossi&64 Zerglings&$\;\;$
MMM&\tabincell{c}{1 Medivac\\2 Marauders\\7 Marines}&\tabincell{c}{1 Medivac\\2 Marauder\\7 Marines}\\
\hline
& & \\[-6pt]
27m\_vs\_30m&27 Marines&30 Marines &$\;\;$
MMM2&\tabincell{c}{1 Medivac\\2 Marauders\\7 Marines}&\tabincell{c}{1 Medivac\\3 Marauder\\8 Marines}\\
\hline
& & \\[-6pt]
25m&25 Marines&25 Marines &$\;\;$
25m\_modified&25 Marines&25 Marines \\
\hline
& & \\[-6pt]
bane\_vs\_bane&\tabincell{c}{4 Banelings\\20 Zerglings}&\tabincell{c}{4 Banelings\\20 Zerglings}&$\;\;$
so\_many\_banelings&7 Zealots&32 Banelings\\
\hline
\end{tabular}
\label{tab:scenario}
\end{table}


\begin{figure*}[htbp]
    \centering
    \subfigure[2s3z]{
        \includegraphics[width=1.45 in]{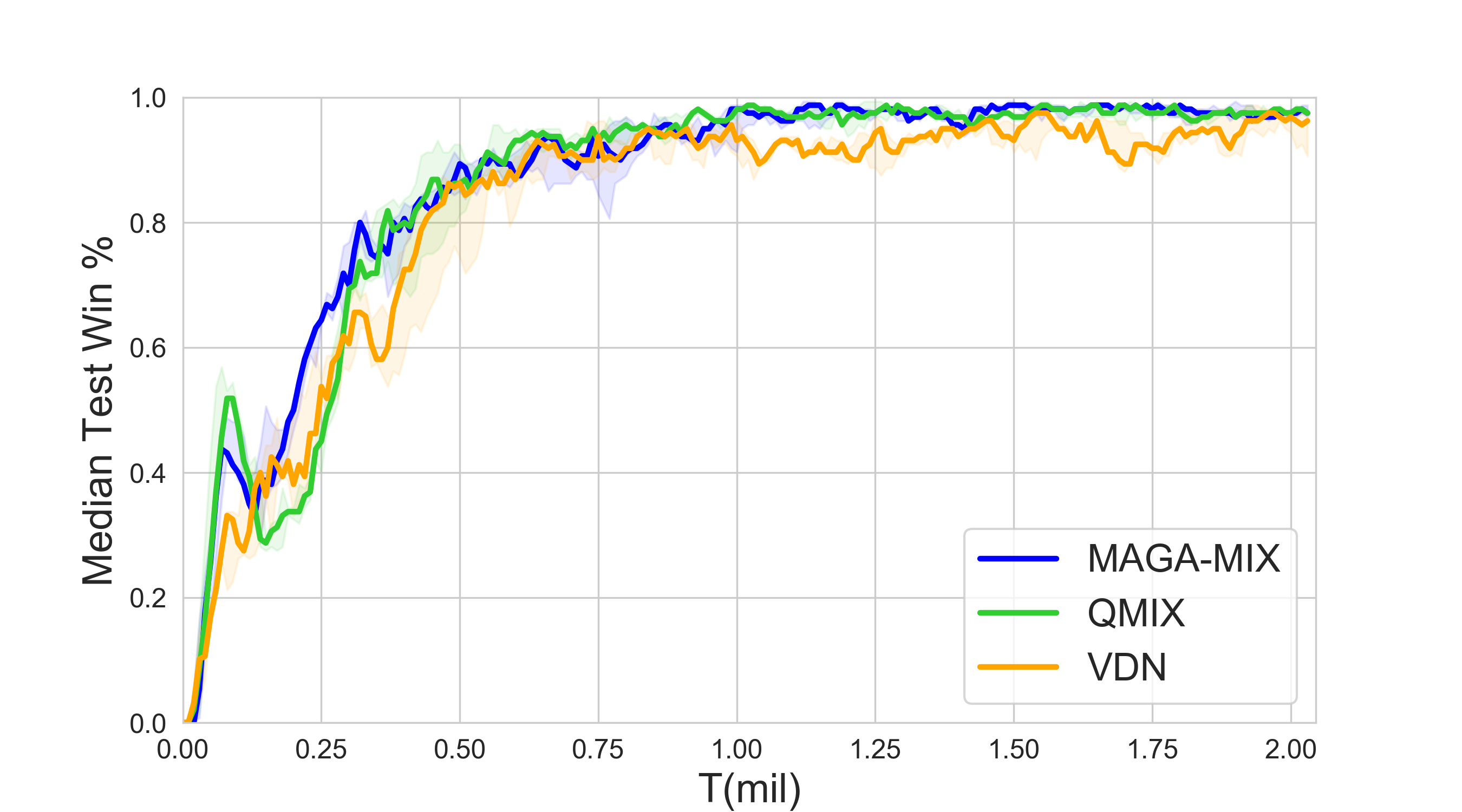}
    }
    \hspace{-0.3 in}
    \subfigure[3s5z]{
        \includegraphics[width=1.45 in]{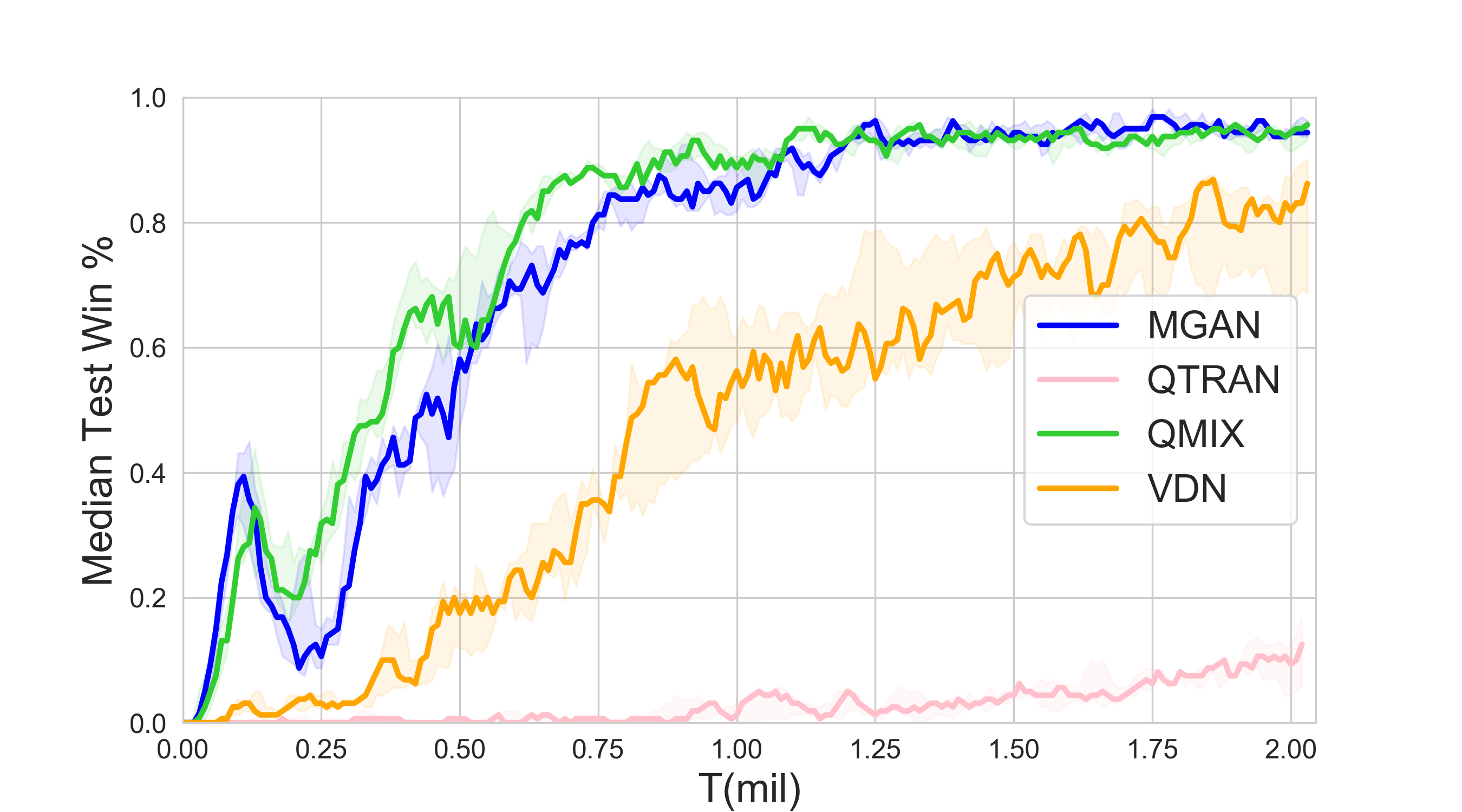}
    }
    \hspace{-0.3 in}
    \subfigure[1c3s5z]{
        \includegraphics[width=1.45 in]{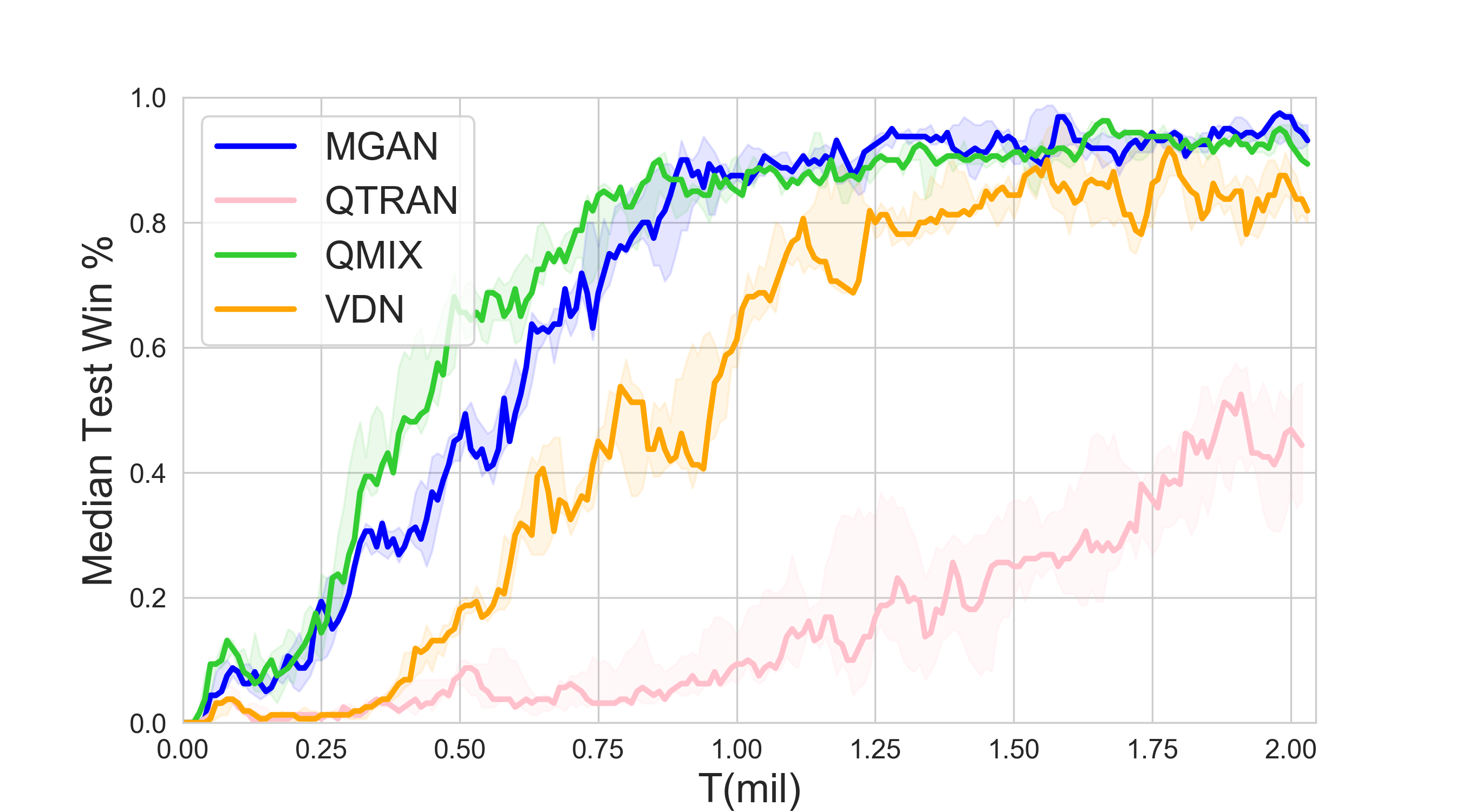}
    }
    \hspace{-0.3 in}
    \subfigure[8m\_vs\_9m]{
        \includegraphics[width=1.45 in]{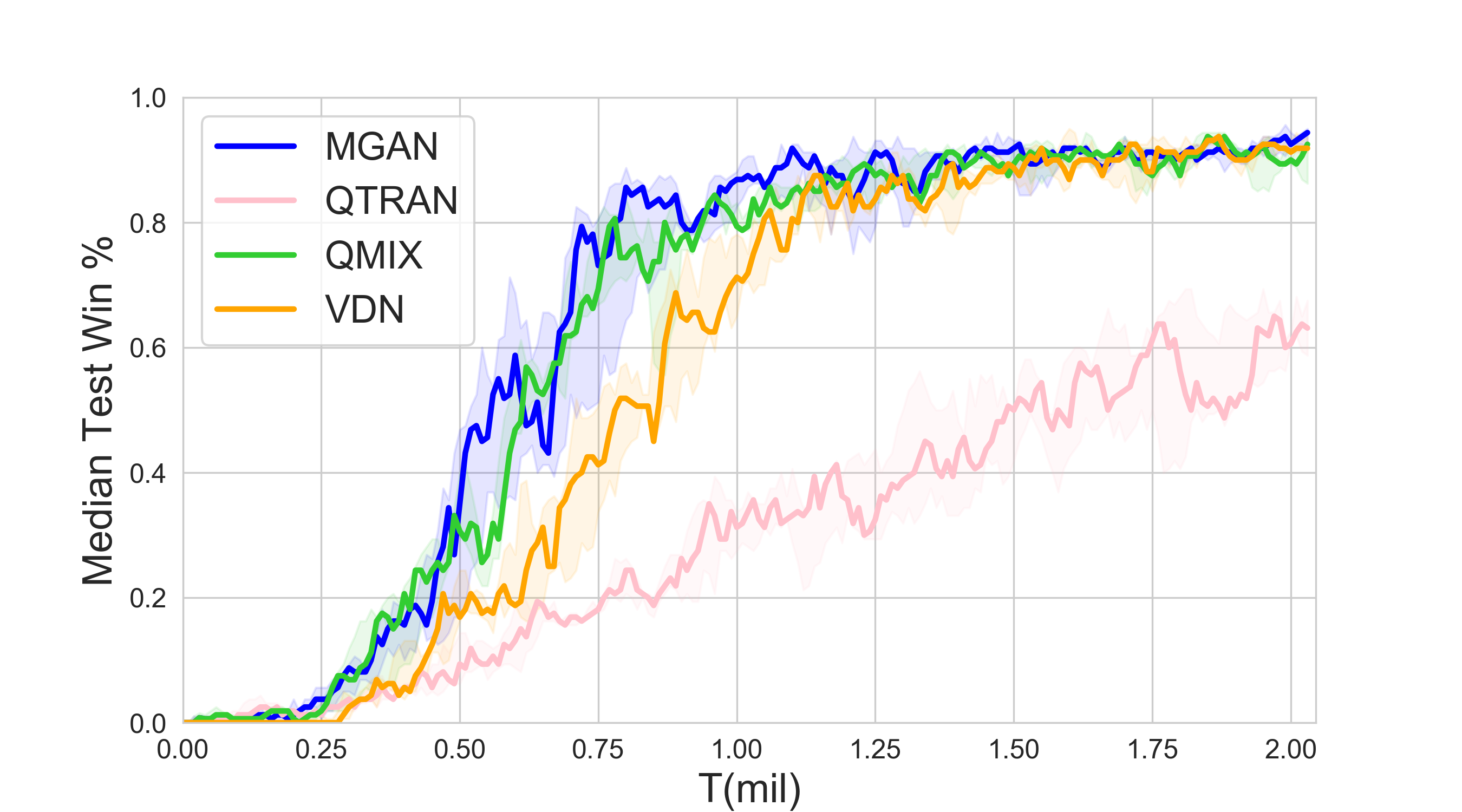}
    }

    \vspace{-0.1 in}
    \subfigure[2c\_vs\_64zg]{
        \includegraphics[width=1.45 in]{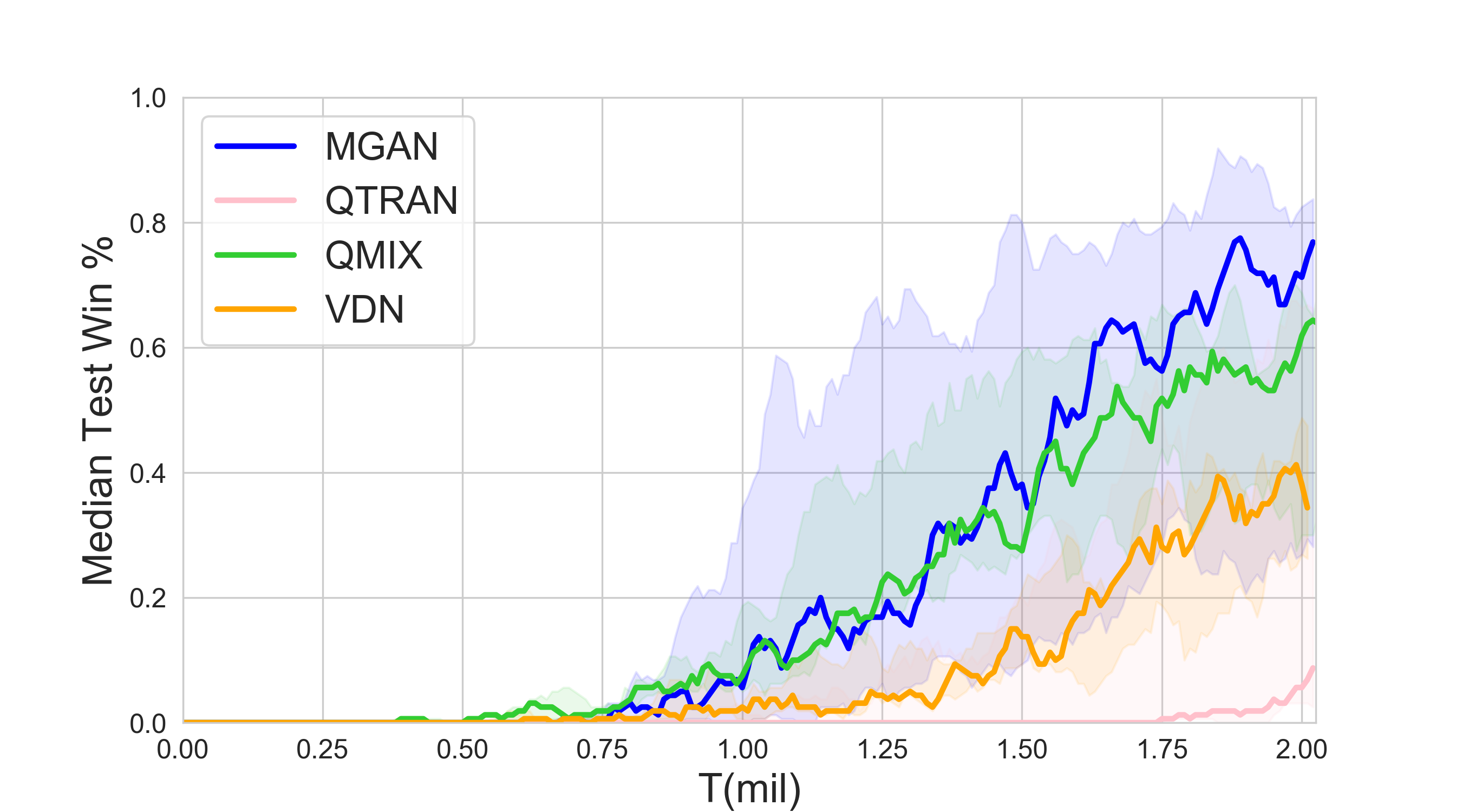}
    }
    \hspace{-0.3 in}
    \subfigure[MMM]{
        \includegraphics[width=1.45 in]{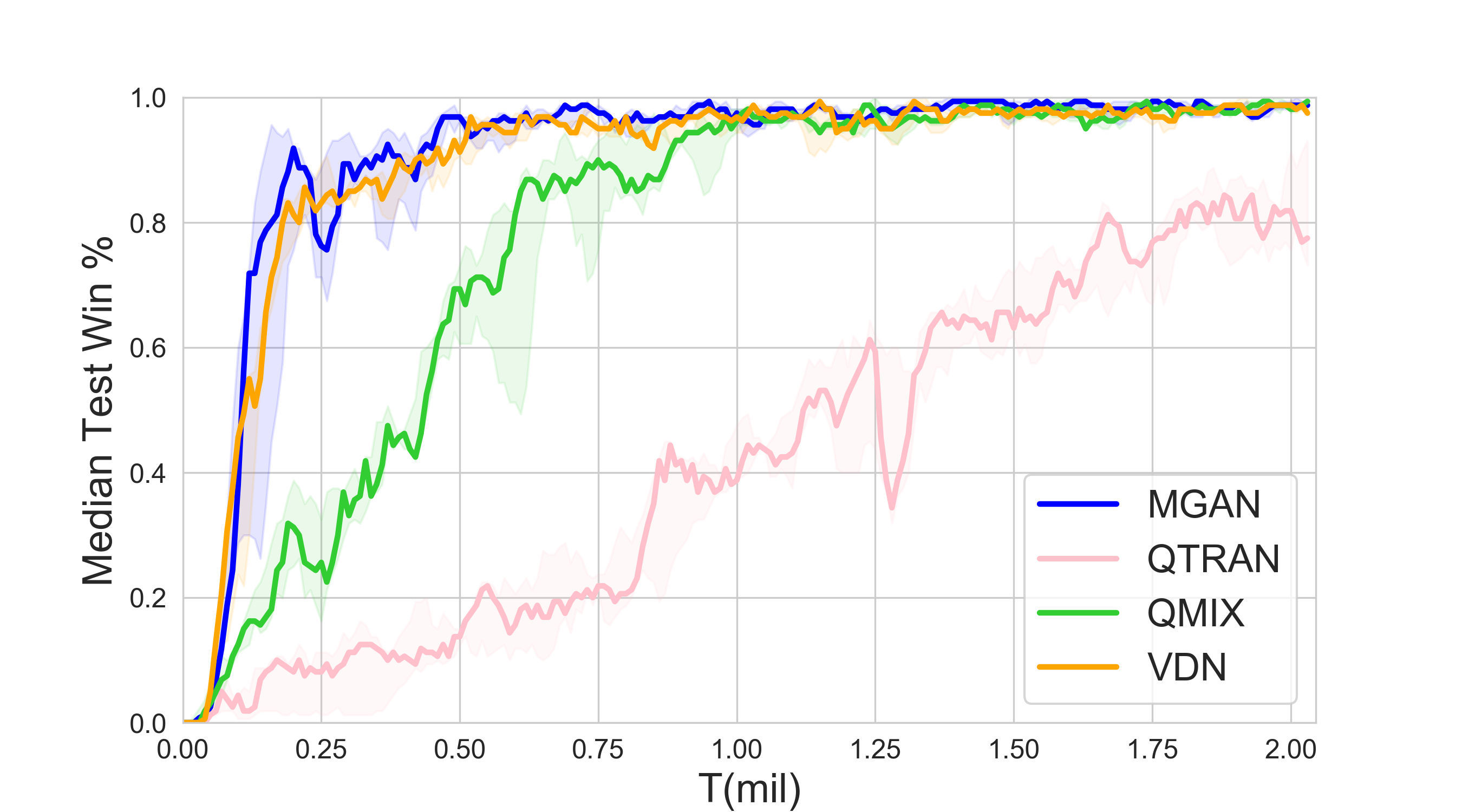}
    }
    \hspace{-0.3 in}
    \subfigure[27m\_vs\_30m]{
        \includegraphics[width=1.45 in]{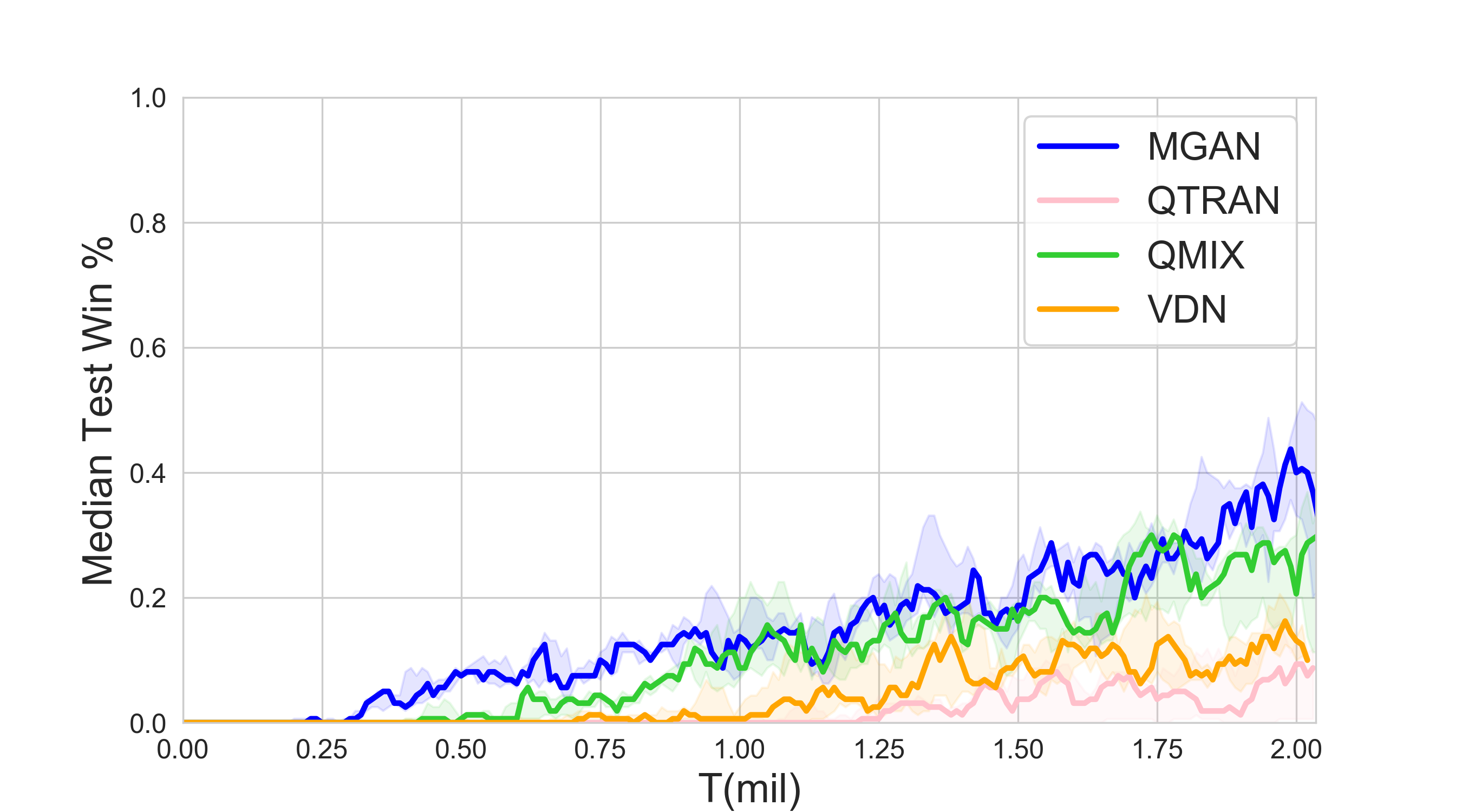}
    }
    \hspace{-0.3 in}
    \subfigure[MMM2]{
        \includegraphics[width= 1.45 in]{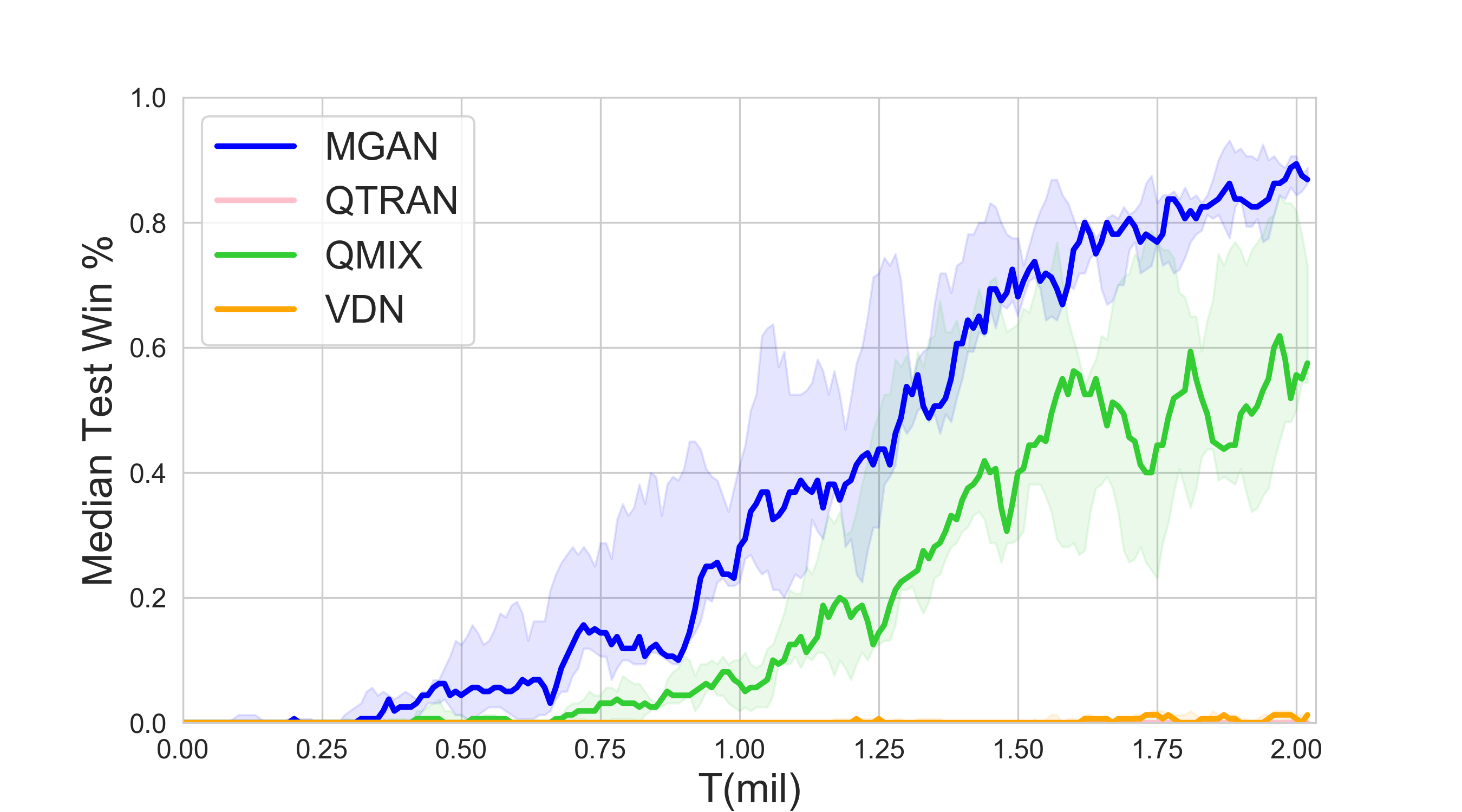}
    }
    
    \vspace{-0.1 in}
    \subfigure[25m]{
        \includegraphics[width=1.45 in]{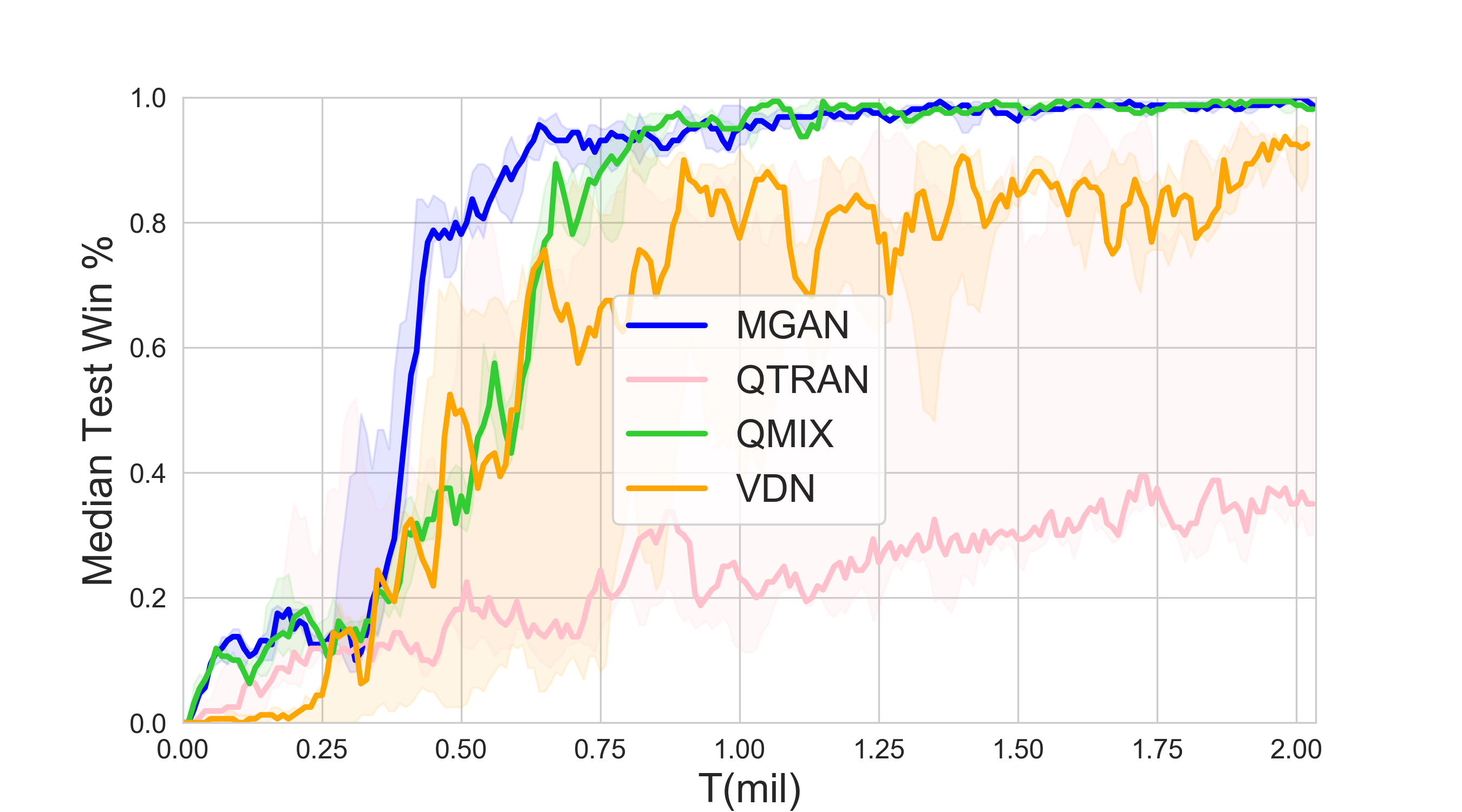}
    }
    \hspace{-0.3 in}
    \subfigure[25m\_modified]{
        \includegraphics[width=1.45 in]{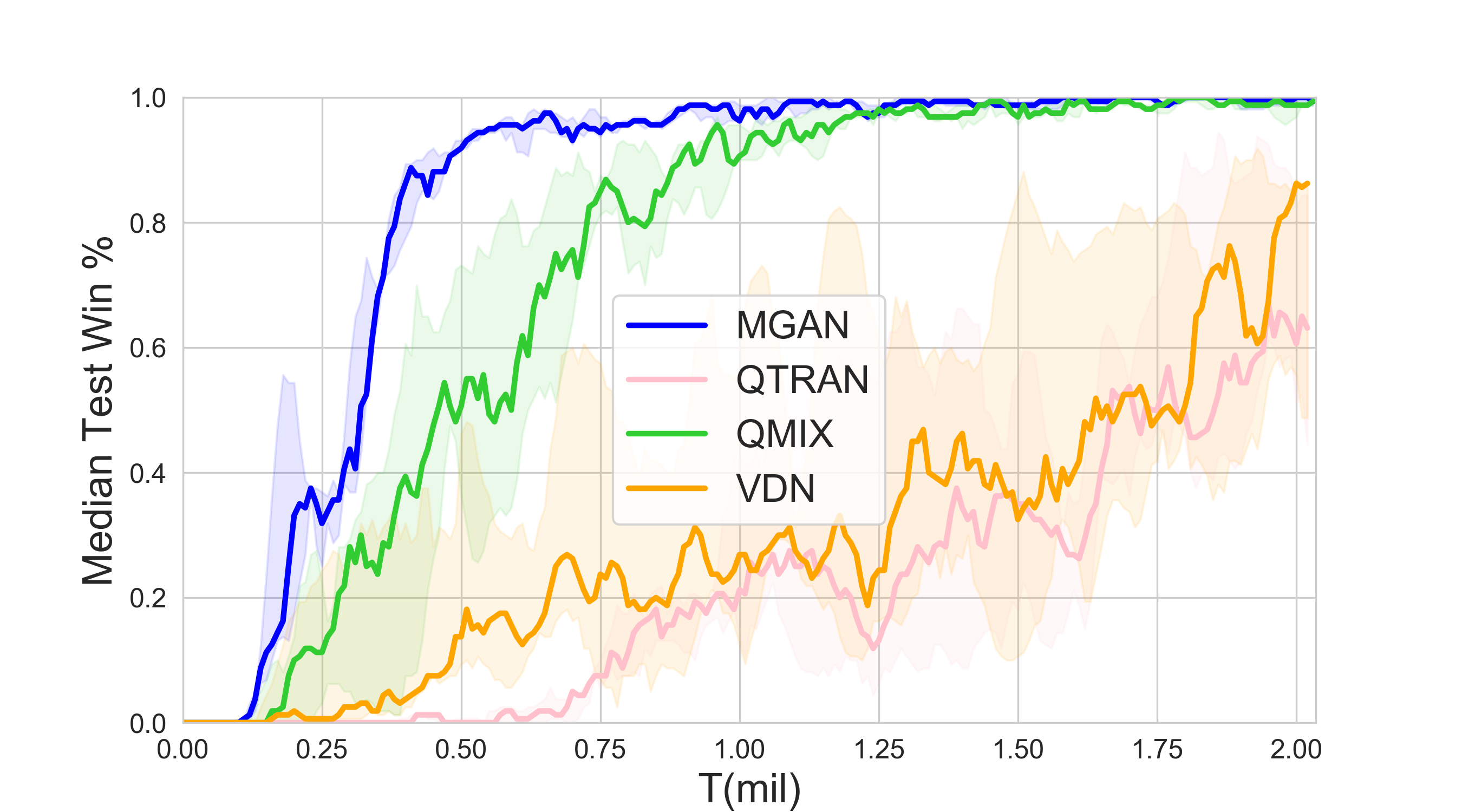}
    }
    \hspace{-0.3 in}
    \subfigure[bane\_vs\_bane]{
        \includegraphics[width=1.45 in]{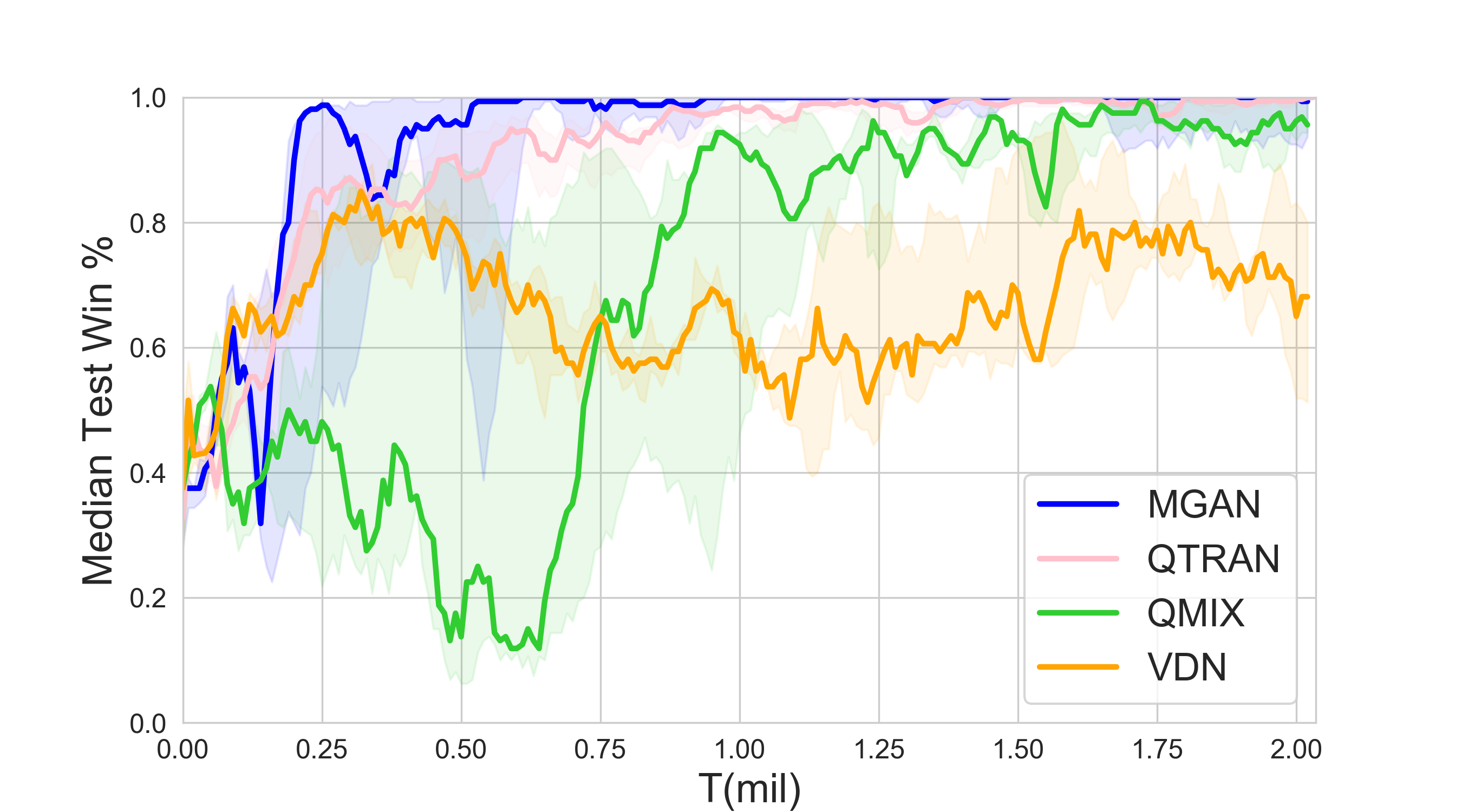}
    }
    \hspace{-0.3 in}
    \subfigure[so\_many\_baneling]{
        \includegraphics[width= 1.45 in]{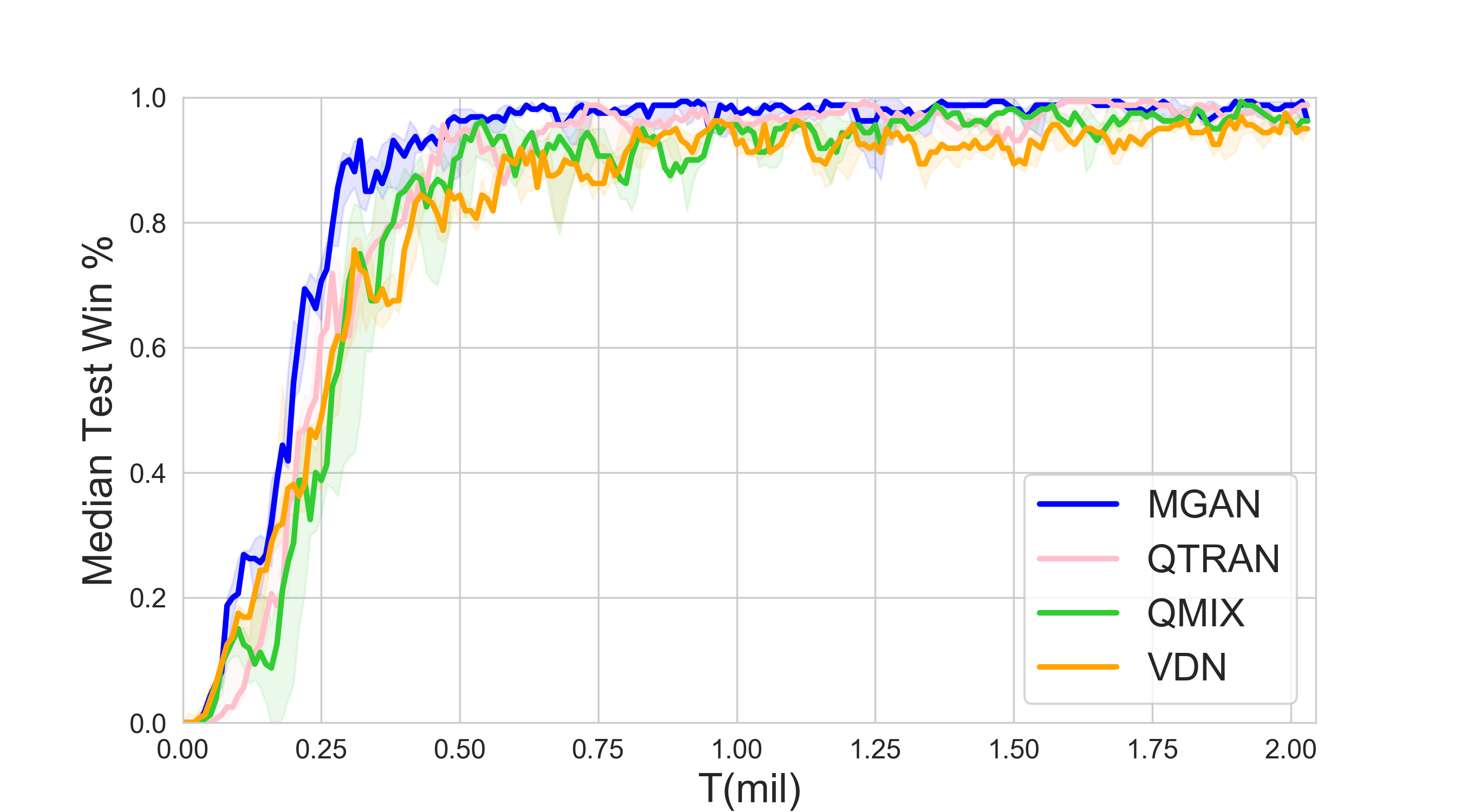}
    }
    \caption{Overall results in different scenarios.}
    \label{fig:result}
\end{figure*}

\section{Experiment}
In this section we will evaluate MGAN and other baselines in the Starcraft II decentralized micromanagement tasks. In addition, to illustrate the representation learning capacity of the graph networks, the visualization of the output of the graph network was performed. We can intuitively understand the motivation of the agents' decision from the output of the graph neural network.

\subsection{Settings}
We use SMAC as the testbed because SMAC is a real-time simulation experiment environment based on Starcraft II. It contains a wealth of micromanagement tasks with varying levels of difficulty. Recently, it has gradually become an important platform for evaluating the coordination capabilities of agents. The scenarios in SMAC include challenges such as asymmetric, heterogeneous, and a large number of agents. We selected more representative scenarios such as \emph{1c3s5z}, \emph{3s5z}, \emph{2c\_vs\_64zg}, \emph{MMM2}, \emph{bane\_vs\_bane} and so on. Besides, in order to be able to more conveniently show MGAN's understanding of the agent in decision-making, we have also introduced a new scenario \emph{25m\_modified}, which is modified on the basis of the \emph{25m} scenario. The distribution of agents in the \emph{25m\_modified} scenario is more dispersed, which makes collaboration more difficult than the original \emph{25m} scenario. The detailed information of all scenarios is shown in the Table~\ref{tab:scenario}.

Our experiment is based on Pymarl\cite{Samvelyan2019TheSM}. We set the hyperparameters of QMIX and VDN to the default in Pymarl. The version of the Starcraft II is 4.6.2(B69232) in our experiments. The feature of each node in the graph network is initialized as its local observation in our proposed MGAN. And according to Equation~(\ref{eq:adjmatrix}), the adjacency matrix $\mathcal{E}$ is given by:
\begin{equation*}
    e_{uv}=\begin{cases}1,&\text{if }u \text{ is alive and } v \text{ is alive}\\0,& \text{otherwise}\end{cases} \qquad  \forall e_{uv} \in \mathcal{E}.
\end{equation*}

The number of graph networks $G$ is set to 4, and the other settings are the same as those of other baselines. We run each experiment 5 times independently to alleviate the effects of accidents and outliers. Depending on the complexity of the experimental scenario, the duration of each experiment ranges from 5 to 14 hours. Experiments are carried out on Nvidia GeForce RTX 3090 graphics cards and Intel(R) Xeon(R) Platinum 8280 CPU. The model is evaluated every 10,000 steps in the experiment, i.e., 32 episodes are run and the win rate is recorded. The agents follow a completely greedy strategy during evaluation.

\subsection{Validation}

Figure~\ref{fig:result} shows the performance results of MGAN and other baselines in different scenarios. The solid line represents the median win ratio of the five experiments. The 25-75\% percentiles of the win ratios are shaded. It can be observed that in some scenarios with a large number of agents, MGAN far exceeds other algorithms in performance. Especially in \emph{bane\_vs\_bane}, MGAN quickly reached convergence. In other scenarios, MGAN is still comparable to other popular algorithms.
\begin{figure*}[htbp]
    \centering
    \subfigure[2nd step on \emph{25m} scenario]{
        \includegraphics[width=0.45\textwidth]{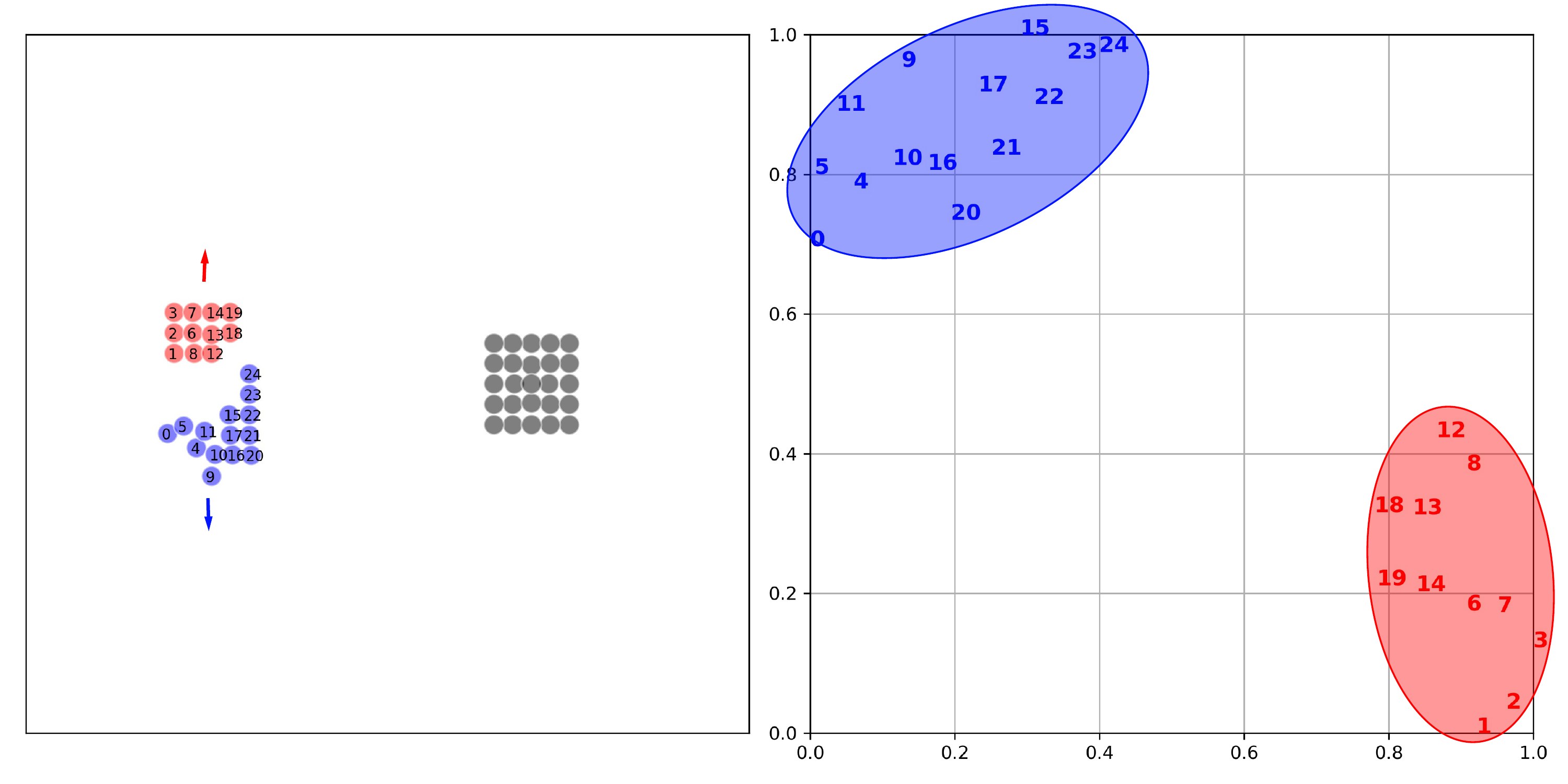}
    }
    \hspace{0.3 in}
    \subfigure[6th step on \emph{25m} scenario]{
        \includegraphics[width=0.45\textwidth]{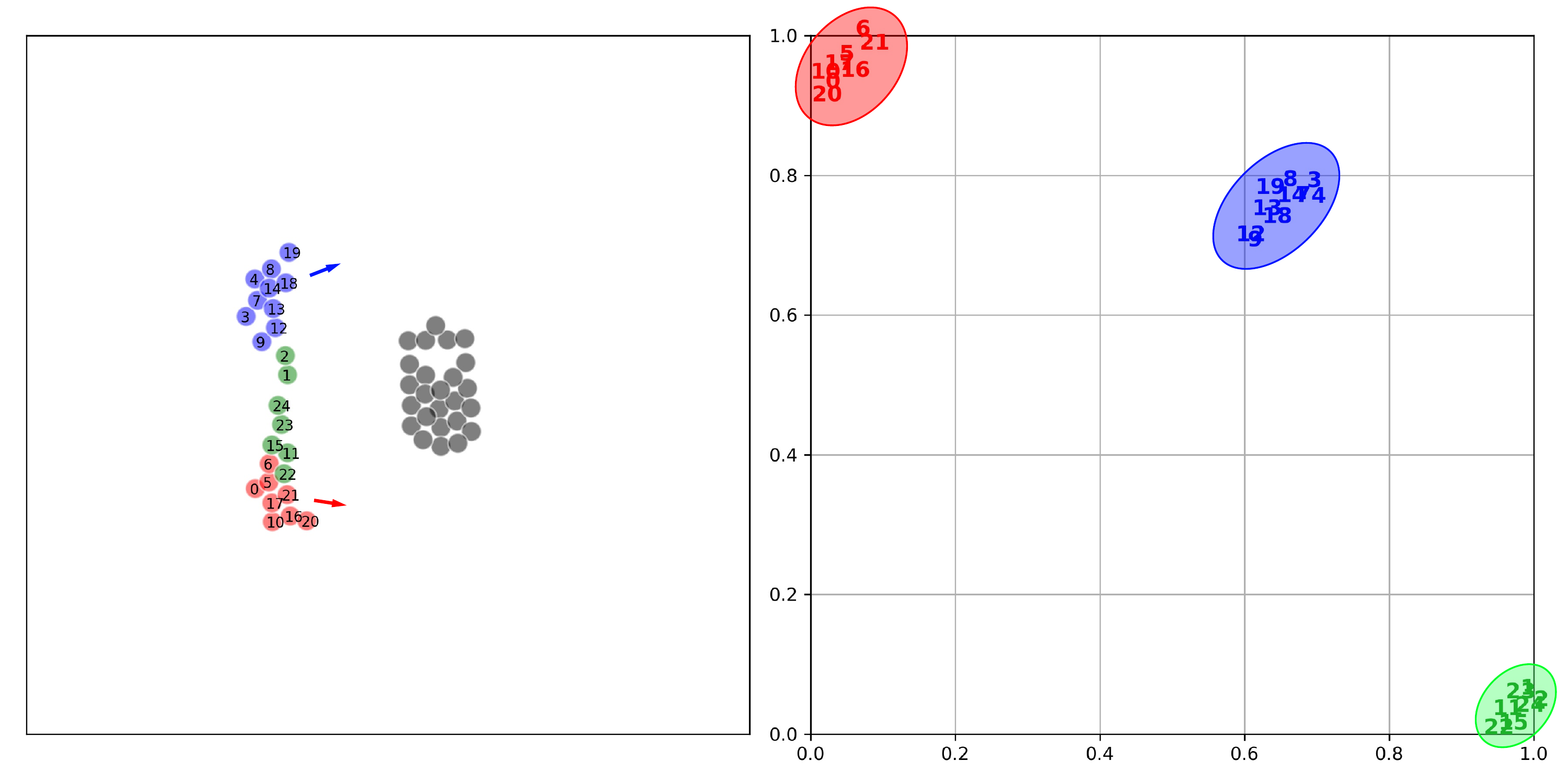}
    }

    \vspace{-0.1 in}
    \subfigure[1st step on \emph{25m\_modified} scenario]{
        \includegraphics[width=0.45\textwidth]{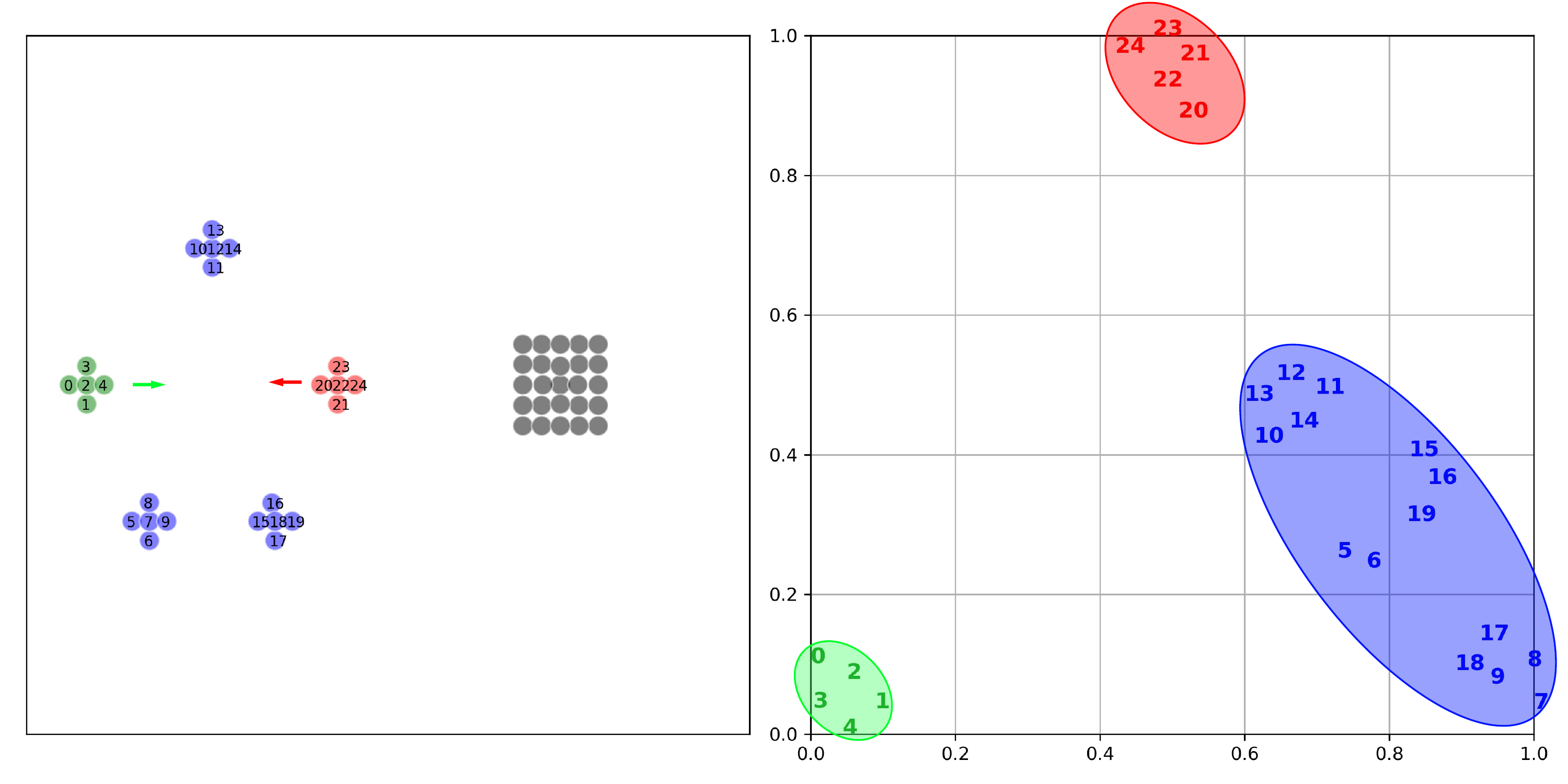}
    }
    \hspace{0.3 in}
    \subfigure[8th step on \emph{25m\_modified} scenario]{
        \includegraphics[width=0.45\textwidth]{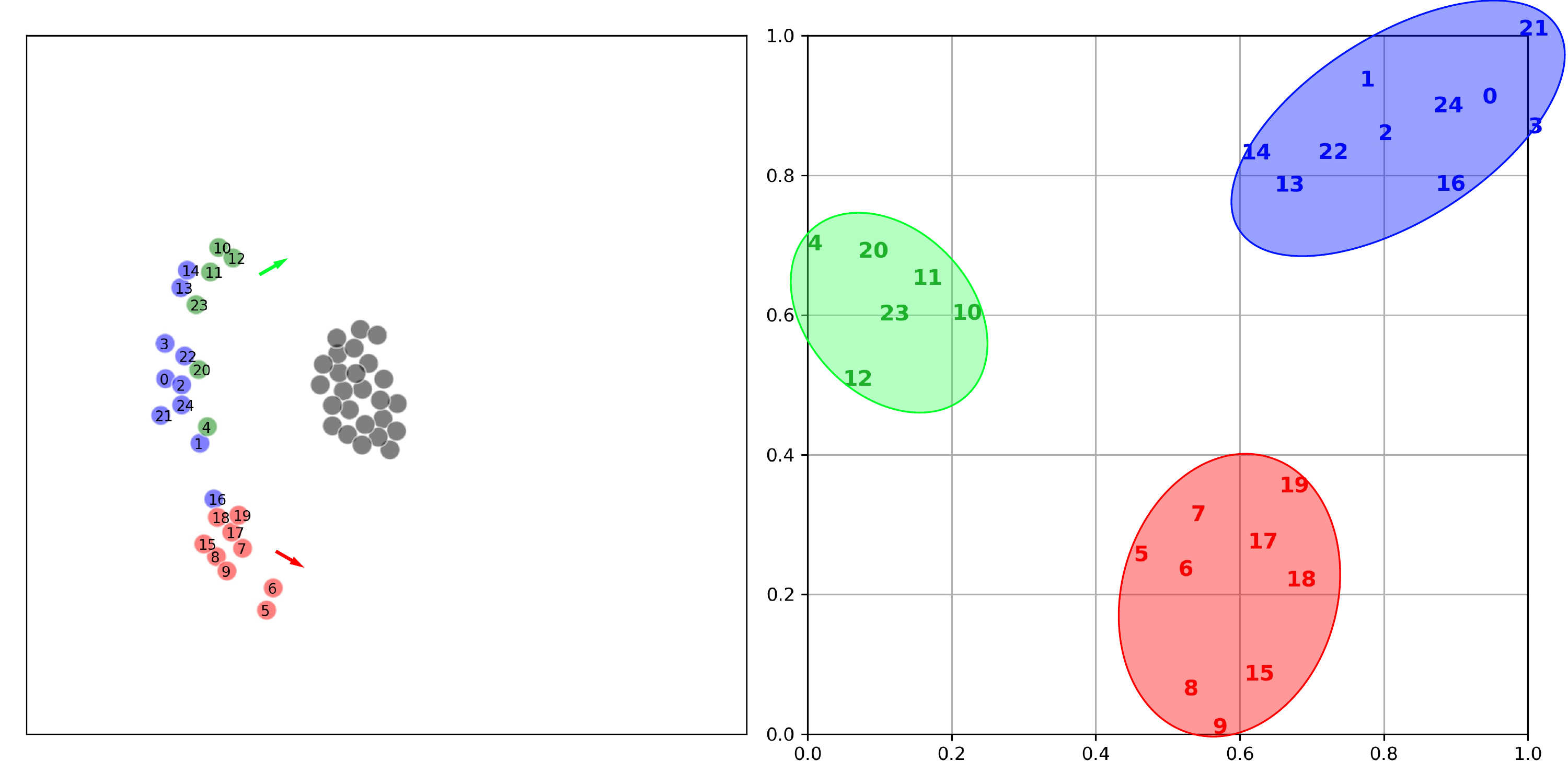}
    }
    \caption{The agents location map at specific step (left) and the corresponding 2D t-SNE embedding of agents' internal states output by one of graph convolutional networks (right). Gray dots in location map represent the enemy agents and color dots denote the agents controlled by MGAN. Each number in 2D t-SNE embedding corresponds to each color dot in the location map one by one.}
\label{fig:tsne}
\end{figure*}

\begin{figure*}[h]
    \centering
    \subfigure[The health values in one episode]{
        \includegraphics[width=0.47\textwidth]{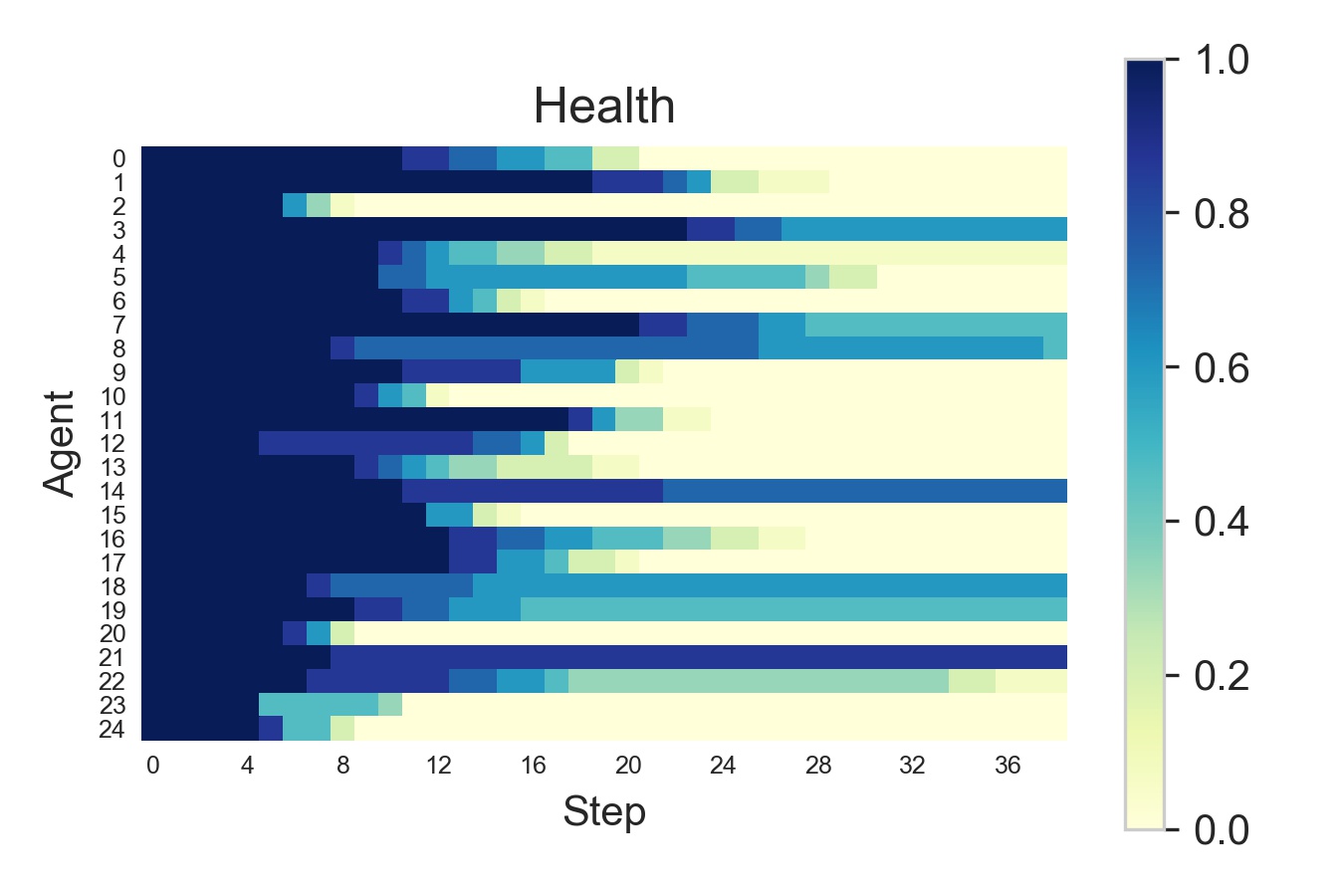}
    }
    \hspace{-0.1 in}
    \subfigure[The weight values in one episode]{
        \includegraphics[width=0.47\textwidth]{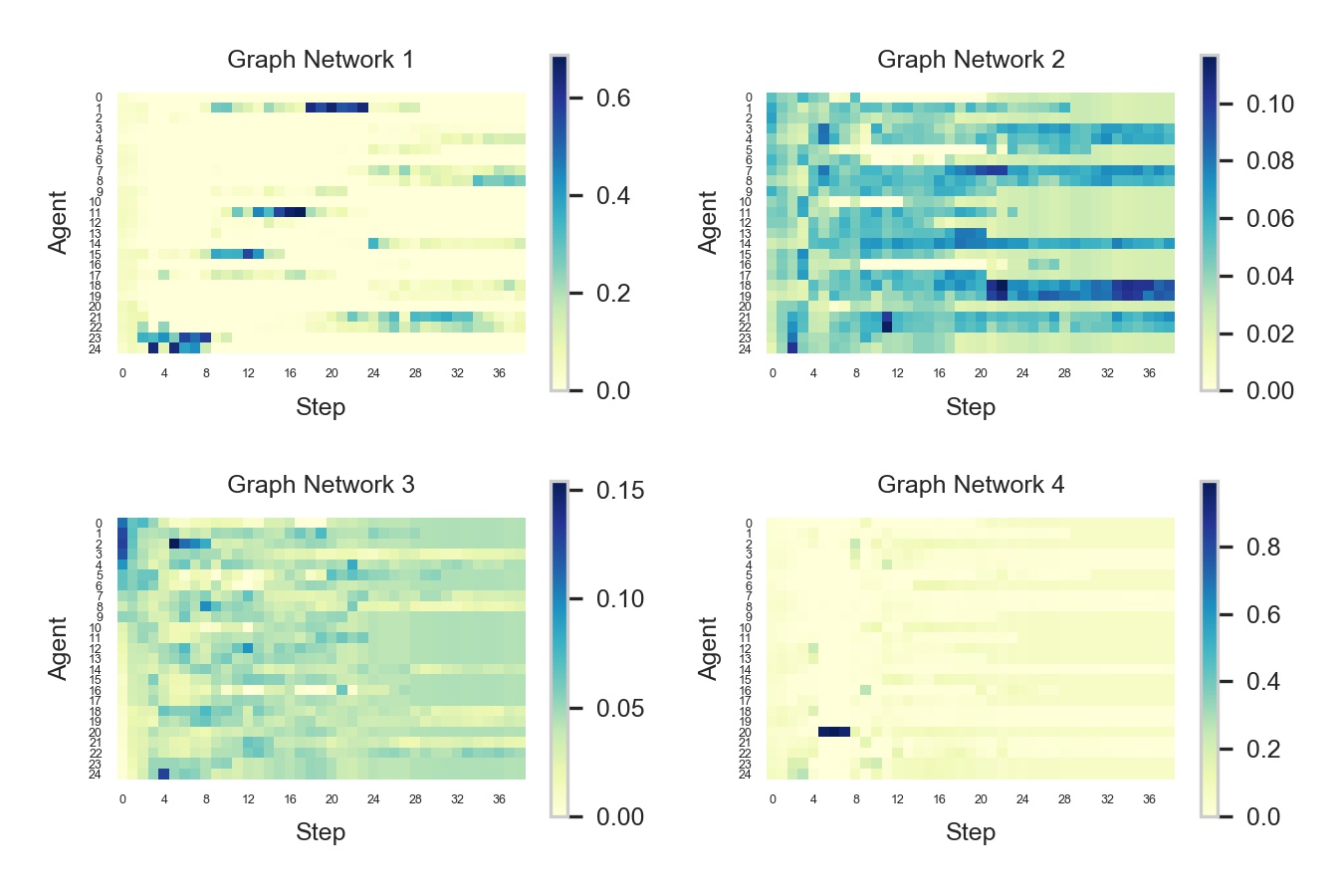}
    }
    \caption{The health values and the weight values on \emph{25m} scenario.}
\label{fig:weights}
\end{figure*}

\begin{table}[h]
\centering
\caption{Median performance of the test win percentage in different scenarios.}
\begin{tabular}{lcccc|lcccc}
\hline
Scenario&MGAN&QTRAN&QMIX&VDN&$\;\;$Scenario&MGAN&QTRAN&QMIX&VDN\\
\hline
& & & & & & & & \\[-6pt]
2s3z&\textbf{98}&93&\textbf{98}&97 &$\;\;$
3s5z&\textbf{97}&13&96&87 \\
1c3s5z&\textbf{96}&53&95&88&$\;\;$
8m\_vs\_9m&\textbf{95}&65&92&92\\
2c\_vs\_64zg&\textbf{77}&9&64&41&$\;\;$
MMM&98&85&\textbf{99}&98 \\
27m\_vs\_30m&\textbf{44}&10&30&16&$\;\;$
MMM2&\textbf{90}&0&62&1\\
25m&\textbf{100}&40&\textbf{100}&94&$\;\;$
25m\_modified&\textbf{100}&67&\textbf{100}&87\\
bane\_vs\_bane&\textbf{100}&\textbf{100}&99&82&$\;\;$
so\_many\_baneling&\textbf{100}&99&99&97\\
\hline
\end{tabular}
\label{tab:result}
\end{table}

Table~\ref{tab:result} shows the median test win rate of different algorithms. As follows from Figure~\ref{fig:result} shown above, it can be seen intuitively that MGAN performs well in hard and super hard scenarios such as \emph{MMM2}, \emph{bane\_vs\_bane} and \emph{27m\_vs\_30m}.

\subsection{Graph Embedding and Weight Analysis}

In order to understand the working principle of MGAN and explore the reasons for its effect improvement, we visualized the embedding vectors output by the graph network and the scalar weights output by the transform layer. We think these two provide an explanatory basis for the agents' actions.

We choose the \emph{25m} and its variant \emph{25m\_modified} scenario with a large number of agents, and show the positions of the agents at each step in the task as a scatter diagram. Meanwhile, t-SNE\cite{Maaten2008VisualizingDU} and MeanShift\cite{Comaniciu2002MeanSA} clustering methods are performed on the graph embedding vector corresponding to each agent in each step, and the corresponding relationship between the position of the agent and the clustering result can be clearly found. This is illustrated in Figure~\ref{fig:tsne}.

In the \emph{25m} scenario, the key to victory is that our agents can form an arc that surrounds the enemy agents. At the beginning of the episode, all agents gathered together. From the results of dimensionality reduction and clustering of embedding vectors, it can be found that the agents are divided into two groups, one group moves upward and the other moves downward. In the middle of the episode, in order to form a relatively concentrated line of fire, the agents was divided into three parts and moved in three directions respectively. In the \emph{25m\_modified} scenario, the agents also need to form the same arc, so the leftmost group of agents needs to move to the right, and the leftmost group of agents needs to move to the left to rendezvous with other agents. And in the middle of the episode, it will still be divided into three parts similar to the \emph{25m} scenario. The finding was quite surprising and suggests that agents in the same subgroup can act together.

For the visualization of the weights, we still use the \emph{25m} scenario for verification. The figure shows the change in the health values of the agents in an episode and the change in the weights of each agent corresponding to the four graph networks. As can be seen from Figure~\ref{fig:weights}, although the values of the weights given by each graph network is not the same, they all have a relationship with the health values of the agents. For example, Graph network 1 believes that agents with drastic changes in health values are the most important ones, while Graph network 2 believes that agents with more health values are the most important. On the contrary, Graph network 3 and Graph network 4 pay more attention to agents whose health values are zero. We guess that this is because these agents cause harm to the enemy and therefore pay more attention.

Through the analysis, we have concluded that the graph network can learn the characteristics of each agent well, and this provides basis for our understanding of the actions of the agents, which improves the interpretability of the motivation of the agents.

\section{Conclusion }
In this paper, we propose a MARL algorithm called MGAN that combines graph network and value-decomposition. From the outcome of our experiments it is possible to conclude that MGAN is comparable to the common baseline, especially in scenarios with a large number of agents. The figures obtained by visualization indicate that the performance improvement is brought about by the graph networks. The findings suggest that this method could also be useful for the works to understand how agents make decisions and what roles they play.

Since MGAN still needs to satisfy the IGM assumption, in our future research we intend to concentrate on how to relax the restrictions of the mixing networks. On the basis of the promising findings presented in this paper, work on the remaining issues is continuing and will be presented in future papers.
\bibliographystyle{splncs04}
\bibliography{reference.bib}

\end{document}